\documentstyle[prd,preprint,tighten,aps,eqsecnum,amssymb,newlfont,epsfig]{revtex}
\newcommand{\kkk}{$K^0 \bar K^0$}

\begin{document}
\begin{title}
\preprint{UWThPh-2001-3\\
January 2001}\\

\center{\textbf{\Huge{Bell inequalities for entangled kaons and their unitary time evolution}}}

\vspace{1.5cm}

\center{\large{Reinhold A. Bertlmann and Beatrix C. Hiesmayr}\footnote{hies@thp.univie.ac.at}\\
\normalsize{{\em Institute for Theoretical Physics, University of Vienna}}\\
{\em Boltzmanngasse 5, A-1090 Vienna, Austria}\\
}
\date{}
\vspace{2.5cm}
\end{title}
\begin{abstract}
We investigate Bell inequalities for neutral kaon systems from $\Phi$ resonance
decay to test local realism versus
quantum mechanics.
We emphasize the unitary time evolution of the states, that means we also include
\emph{all} decay product states, in contrast to other
authors. Only this guarantees the use
of the complete Hilbert space. We develop a general formalism for Bell inequalities
including both arbitrary ``quasi spin'' states and different times; finally
we analyze Wigner-type inequalities.
They contain an additional term, a correction function $h$, as compared to the
spin $1/2$ or photon case, which changes considerably the possibility of
quantum mechanics to violate the Bell inequality. Examples for special ``quasi
spin''
states are given, especially those which are sensitive to the $CP$ parameters
$\varepsilon$ and $\varepsilon'$.
\end{abstract}

\vspace{2cm}
PACS number(s): 13.25.Es, 03.65.Bz, 14.40.Aq
entangled states

\newpage

\section{Introduction}

Already Schr\" odinger \cite{Schrodinger} in 1935 pointed to the peculiar
features of what he called entangled states (``verschr\" ankte Zust\" ande''
in his original words). It was Einstein, Podolsky and Rosen (EPR) who aimed
in their famous paper \cite{EPR} to show the incompleteness of quantum mechanics
(QM) by considering a quantum system of two particles. Also Furry \cite{Furry}
emphasized, inspired by EPR and Schr\" odinger, the differences between the
predictions of QM of non-factorizable systems and models with
spontaneous factorization.

Much later, in 1964, this subject was brought up again by John S. Bell \cite{bell}
who re-analyzed the ``EPR paradox''. He discovered via an inequality that the
predictions of QM differ from those of (all) local realistic theories (LRT);
inequalities of this type are now named quite generally ``Bell inequalities''.

It is the nonlocality, the ``spooky action at a distance'', which is the basic feature of quantum
physics and is so contrary to
our intuition or more precise, the nonlocal correlations
between the spatially separated EPR pair, which occur due to the quantum entanglement.

Many beautiful experiments have been carried out over the years (see e.g. Refs.
\cite{FreedmanClauser,FryThompson,aspect,zeilinger})
by using the entanglement of the polarization of two photons;
all confirm impressively this very peculiar quantum feature.

The nonlocality does not conflict with Einstein's relativity, so it cannot be used
for superluminal communication, nevertheless, it is the basis for new physics,
like quantum cryptography \cite{Ekert,DeutschEkert,Hughes,GisinGroup} and
quantum teleportation \cite{Bennett,ZeilingerTele}, and it triggered a new technology:
quantum information \cite{ZeilingerInfo1,ZeilingerInfo2}.

Of course, it is of great interest to test the EPR-Bell correlations also for
massive systems in particle physics. Already in 1960 Lee and Yang \cite{lee}
and several other authors \cite{Inglis,Day,Lipkin}
emphasized the EPR-like features of a \kkk  pair in a $J^{PC}=1^{--}$ state.
Indeed many authors \cite{eberhard,eberhard2,bigi,domenico,Bramon,AncoBramon} suggested to
investigate the \kkk  pairs which are produced at the $\Phi$ resonance -- for instance in the
$e^+ e^-$-machine DA$\Phi$NE at Frascati. And the non separability of the neutral kaon system
-- created in $p\bar{p}$-collisions -- has been analyzed by the authors
\cite{six,CPLEAR-EPR,BGH}.

Similar systems are the entangled $B^0 \bar B^0$ pairs produced at the $\Upsilon$(4S)
resonance (see e.g., Refs. \cite{BG1,Dass,BG2,datta,selleribmeson,BG3}),
which we do not consider here.

Specific realistic theories have been constructed \cite{selleri,SelleriBook,six2},
which describe the \kkk pairs, as tests versus quantum mechanics.
However, the general test of QM versus LRT relies on Bell inequalities, where
the different kaon detection times play the role of the different angles in the
photon or spin $1/2$ case. On the other hand, also the free choice
of the kaon ``quasi spin'' state is of importance. Furthermore an interesting
feature of kaons is the $CP$ violation and indeed it turns out
that Bell inequalities imply bounds on the physical $CP$ violation
parameters $\varepsilon$ and $\varepsilon'$. In this connection also
a bound on the degree of decoherence of the wavefunction can be
found \cite{trixi2}, which turns out to be very strong for a distinction of QM versus LRT.

The important difference of the kaon systems as compared to
photons is their decay. Focusing, therefore, just on some particular
``quasi spin'' states and not accounting for the decay states
restricts the investigation to a subset of the total Hilbert space
and will limit the validity of the physical theories.

Therefore we allow in our work for the freedom of choosing arbitrary ``quasi spin''
states and we emphasize the importance of including {\it all} decay product states
into the BI, in contrast to other authors, so we use a unitary
time evolution. Only this guarantees the use of the complete
Hilbert space. It may very well happen that in a particular subspace QM
violates indeed the BI for certain times $t>0$ , and thus contradicts with
the assumptions of reality and locality, but in the total Hilbert
space the violation will disappear. We show cases where this will happen.
Note, that for entangled spin $1/2$ particles or photon systems all operations
are already defined on the total Hilbert space, since the photon does not decay and its
polarization is conserved, whereas in the kaon systems strangeness
is not conserved due to the weak interactions.

The paper is organized as follows. In Section $2$ we give an introduction to neutral
kaons and explain the ``quasi spin'' picture. In Section $3$ the unitary time
evolution is worked out in detail and how one has to calculate the
probabilities in quantum mechanics. In Section $4$ we review briefly the Bell inequalities for
spin $1/2$ particles. Our main part is contained in Section $5$, there
we derive the generalized Bell inequalities for entangled kaons and analyze three
different examples which can be found in the literature. Section $6$ summarizes our
results and the conclusions are drawn. Finally, some useful formulae
can be found in the Appendix.

\section{Neutral $K$-mesons}

Let us start with a discussion of the properties of the neutral
kaons, which we need in the following.
The neutral $K$-mesons are characterized by their strangeness quantum number $S$
\begin{eqnarray}
S|K^0\rangle \; &=& + |K^0\rangle\nonumber\\
S|\bar K^0\rangle \; &=& - |\bar K^0\rangle .
\end{eqnarray}
As the $K$-mesons are pseudoscalars their parity $P$ is minus and charge conjugation
$C$ transforms $K^0$ and $\bar K^0$ into each other so that we have for the combined
transformation $CP$ (in our choice of phases)
\begin{eqnarray}
CP|K^0\rangle \; &=& - |\bar K^0\rangle\nonumber\\
CP|\bar K^0\rangle \; &=& - |K^0\rangle .
\end{eqnarray}
From this follows that the orthogonal linear combinations
\begin{eqnarray}
|K_1^0\rangle \; &=& \frac{1}{\sqrt{2}}\big\lbrace |K^0\rangle-
|\bar K^0\rangle \big\rbrace\nonumber\\
|K_2^0\rangle \; &=& \frac{1}{\sqrt{2}}\big\lbrace |K^0\rangle+
|\bar K^0\rangle \big\rbrace
\end{eqnarray}
are eigenstates of $CP$
\begin{eqnarray}
CP|K_1^0\rangle \; &=& + |K_1^0\rangle\nonumber\\
CP|K_2^0\rangle \; &=& - |K_2^0\rangle ,
\end{eqnarray}
a quantum number conserved in strong interactions.

Due to weak interactions, which are $CP$ violating, the kaons decay and the physical
states, having the mass $m_S$ and $m_L$, are the short and long lived states
\begin{eqnarray}\label{kaonSL}
|K_S\rangle \; &=& \frac{1}{N}\big\lbrace p |K^0\rangle-q
|\bar K^0\rangle \big\rbrace\nonumber\\
|K_L\rangle \; &=& \frac{1}{N}\big\lbrace p |K^0\rangle+q
|\bar K^0\rangle \big\rbrace
\end{eqnarray}
with $p=1+\varepsilon$, $q=1-\varepsilon$, $N^2=|p|^2+|q|^2$ and $\varepsilon$ being
the complex $CP$ violating parameter ($CPT$ invariance is assumed; thus the short and long
lived states contain the same $CP$ violating parameter
$\varepsilon_S=\varepsilon_L=\varepsilon$). They are eigenstates of the
non-Hermitian ``effective mass'' Hamiltonian
\begin{equation}\label{hamiltonian}
H \, = \, M - \frac{i}{2} \Gamma
\end{equation}
satisfying
\begin{equation}
H |K_{S,L}\rangle \; = \; \lambda_{S,L} |K_{S,L}\rangle
\end{equation}
with
\begin{equation}
\lambda_{S,L} \, = \, m_{S,L} - \frac{i}{2} \Gamma_{S,L} \,.
\end{equation}

Both mesons $K^0$ and $\bar K^0$ have transitions to common states
(due to $CP$ violation) therefore they mix, that means they oscillate
between $K^0$ and $\bar K^0$ before decaying. Since the decaying states
evolve --- according to the Wigner-Weisskopf approximation --- exponentially in time
\begin{equation}
| K_{S,L} (t)\rangle \; = \; e^{-i \lambda_{S,L} t} | K_{S,L} \rangle \,,
\end{equation}
the subsequent time evolution for $K^0$ and $\bar K^0$ is given by
\begin{eqnarray}
| K^0(t) \rangle  &=&
g_{+}(t) | K^0 \rangle  + \frac{q}{p} g_{-}(t) | \bar K^0 \rangle
\nonumber\\
| \bar K^0(t) \rangle  &=&
\frac{p}{q} g_{-}(t) | K^0 \rangle + g_{+}(t) | \bar K^0 \rangle
\end{eqnarray}
with
\begin{equation}\label{g+-}
g_{\pm}(t) \, = \, \frac{1}{2} \left[ \pm e^{-i \lambda_S t} + e^{-i \lambda_L t}
\right] \,.
\end{equation}
Supposing that at $t=0$ a $K^0$ beam is produced, e.g. by strong
interactions, then the probability for finding a $K^0$ or $\bar K^0$
in the beam is calculated by
\begin{eqnarray}
\left| \langle K^0 | K^0(t) \rangle \right|^2 &=& \frac{1}{4} \frac{|q|^2}{|p|^2}
\big\lbrace e^{-\Gamma_S t} + e^{-\Gamma_L t} + 2 \, e^{-\Gamma t}
\cos(\Delta m t)\big\rbrace\nonumber\\
\left| \langle \bar K^0 | K^0(t) \rangle \right|^2 &=& \frac{1}{4} \frac{|q|^2}{|p|^2}
\big\lbrace e^{-\Gamma_S t} + e^{-\Gamma_L t} - 2 \, e^{-\Gamma t}
\cos(\Delta m t)\big\rbrace \, ,
\end{eqnarray}
with $\Delta m = m_L-m_S$ and $\Gamma = \frac{1}{2}(\Gamma_L+\Gamma_S)$.
The $K^0$ beam oscillates with frequency $\Delta m / 2\pi$, the
oscillation being clearly visible at times of the order of a few $\tau_S$,
before all $K_S$ have died out leaving only the $K_L$ in the beam.
So in a beam which contains only $K_0$ mesons at the time $t=0$
the $\bar K_0$ will appear far from the production source
through its presence in the $K_L$ meson with equal probability
as the $K_0$ meson. A similar feature occurs when starting with a
$\bar K^0$ beam.\\

In comparison with spin $1/2$ particles, or with photons having
the polarization directions vertical and horizontal, it is
especially useful to work with the ``quasi-spin'' picture for kaons
introduced by Lee and Wu \cite{LeeWu} and Lipkin \cite{Lipkin}.
The two states $| K^0 \rangle$ and $| \bar K^0 \rangle$ are
regarded as the quasi-spin states up $|\Uparrow\rangle$ and down
$|\Downarrow\rangle$ and the operators acting in this quasi-spin
space are expressed by Pauli matrices. So the strangeness operator
$S$ can be identified with the Pauli matrix $\sigma_3$, the $CP$
operator with ($-\sigma_1$) and $CP$ violation is proportional to
$\sigma_2$. In fact, the Hamiltonian (\ref{hamiltonian}) can be written as
\begin{equation}
H \, = \, a\cdot \mathbf{1} + \vec b \cdot \vec \sigma
\end{equation}
with
\begin{eqnarray}
b_1 = b \cos \alpha, \quad b_2 = b \sin \alpha, \quad b_3 = 0\nonumber\\
a = \frac{1}{2}(\lambda_L + \lambda_S), \quad
b = \frac{1}{2}(\lambda_L - \lambda_S)
\end{eqnarray}
($b_3 = 0$ due to $CPT$ invariance), and the phase $\alpha$ is
related to the $CP$ parameter $\varepsilon$ by
\begin{equation}
e^{i\alpha} \, = \, \frac{1-\varepsilon}{1+\varepsilon} \, .
\end{equation}

Now, what we are actually interested in are entangled states of
\kkk pairs, in analogy to the entangled spin up and down pairs, or
photon pairs. Such states are produced by $e^+ e^-$-machines
through the reaction $e^+ e^- \to \Phi \to K^0 \bar K^0$, in
particular at DA$\Phi$NE, or they are produced in $p\bar
p$-collisions like, e.g. at LEAR. There a \kkk pair is created in
a $J^{PC}=1^{--}$ quantum state and thus antisymmetric under $C$
and $P$, and is described at the time $t=0$ by the entangled state
\begin{eqnarray}\label{entangledK0}
| \psi (t=0) \rangle &=&\frac{1}{\sqrt{2}}
\left\{ | K^0 \rangle_l \otimes\! | \bar K^0 \rangle _r -
| \bar K^0 \rangle _l \otimes | K^0 \rangle _r \right\},
\end{eqnarray}
which can be rewritten in the $K_S K_L$-basis
\begin{eqnarray}
| \psi (t=0) \rangle&=&
 \frac{N_{SL}}{\sqrt{2}}\left\{ | K_S \rangle_l \otimes\! | K_L \rangle _r -
| K_L \rangle _l \otimes | K_S \rangle _r \right\}
\end{eqnarray}

with $N_{SL}=\frac{N^2}{2pq}$. Then the neutral kaons fly apart and will be detected
on the left ($l$)
and right ($r$) side of the source. Of course, during their propagation
the \kkk oscillate and $K_S, K_L$ decays will take place. This is
an important difference to the case of spin $1/2$ particles or
photons.

\section{Time evolution - unitarity}

Now let us discuss more closely the time evolution of the kaon
states \cite{BellSteinberger} . At any instant $t$ the state
$| K^0(t) \rangle$ decays to a specific final state $| f \rangle$
with a probability proportional to the absolute squared of the
transition matrix element. Because of unitarity of the time evolution
the norm of the total state must be conserved. This means that the decrease
in the norm of the state $| K^0(t) \rangle$ must be compensated by the
increase in the norm of the final states.

So starting at $t=0$ with a $K^0$ meson the state we have to
consider for a complete $t$-evolution is given by
\begin{eqnarray}
|K^0\rangle\;\stackrel{\longrightarrow}\;\;a(t)|K^0\rangle+b(t)|\bar K^0\rangle+\sum_f
c_f(t) |f\rangle
\end{eqnarray}
with
\begin{eqnarray}
a(t)=g_+(t)\qquad\textrm{and}\qquad b(t)=\frac{q}{p} g_-(t)
\end{eqnarray}
and the functions $g_{\pm}(t)$ are defined in Eq.(\ref{g+-}).
Denoting the amplitudes of the decays of the $K^0, \bar K^0 \,$ to a specific
final state $f$ by
\begin{eqnarray}
\mathcal{A}(K^0\longrightarrow f)\equiv \mathcal{A}_f\qquad\textrm{and}\qquad
\mathcal{A}(\bar K^0 \longrightarrow f)\equiv \bar{\mathcal{A}}_f
\end{eqnarray}
we have
\begin{eqnarray}
\frac{d}{dt}|c_f(t)|^2=|a(t) \mathcal{A}_f+b(t) \bar{\mathcal{A}}_f|^2
\end{eqnarray}
and for the probability of the decay $K_0 \to f$ at a certain time $\tau$
\begin{eqnarray}
P_{K^0\longrightarrow f}(\tau)=\int_0^\tau \frac{d}{dt}|c_f(t)|^2 dt \, .
\end{eqnarray}

Since the state $| K^0(t) \rangle$ evolves according to a
Schr\"odinger equation with ``effective mass'' Hamiltonian (\ref{hamiltonian})
the decay amplitudes are related to the $\Gamma$ matrix by
\begin{eqnarray}
\Gamma_{11}=\sum_f |\mathcal{A}_f|^2,\quad \Gamma_{22}=\sum_f
|\bar\mathcal{A}_f|^2,\quad \Gamma_{12}=\sum_f \mathcal{A}_f^* \bar \mathcal{A}_f \; .
\end{eqnarray}
These are the Bell-Steinberger unitarity relations \cite{BellSteinberger};
they are a consequence of probability conservation, and play an important
role.\\

For our purpose the formalism used by Ghirardi, Grassi and Weber \cite{ghirardi91}
is quite convenient, and we generalize it to arbitrary quasi spin states.
So we describe the complete evolution of the mass eigenstates by a unitary
operator $U(t,0)$ whose effect can be written as
\begin{eqnarray}\label{timeevolution}
U(t,0)\; |K_{S,L}\rangle &=& e^{-i \lambda_{S,L} t}\;|K_{S,L}\rangle +
|\Omega_{S,L}(t)\rangle
\end{eqnarray}
where $|\Omega_{S,L}(t)\rangle$ denotes the state of all decay products.
For the transition amplitudes of the decay product states we then have
\begin{eqnarray}
\langle \Omega_S(t)|\Omega_S(t)\rangle&=& 1-e^{-\Gamma_S t}\\
\langle \Omega_L(t)|\Omega_L(t)\rangle&=& 1-e^{-\Gamma_L t}\\
\langle \Omega_L(t)|\Omega_S(t)\rangle&=&\langle K_L|K_S\rangle
(1-e^{i \Delta m t}e^{-\Gamma t})\\
\langle K_{S,L}|\Omega_S(t)\rangle&=&\langle K_{S,L}|\Omega_L(t)\rangle=0 \;.
\end{eqnarray}

Note that the mass eigenstates (\ref{kaonSL}) are normalized but
due to $CP$ violation not orthogonal

\begin{eqnarray}
\langle K_L|K_S\rangle = \frac{2 Re\{\varepsilon\}}{1+|\varepsilon|^2} =: \delta \;.
\end{eqnarray}

Now we consider entangled states of kaon pairs, and we start at time $t=0$ from
the entangled state (\ref{entangledK0}) given in the $K_S K_L$ basis choice
\begin{equation}\label{entangledKS}
|\psi(t=0)\rangle \;=\; \frac{N^2}{2\sqrt{2} p q}\big\lbrace
|K_S\rangle_l \otimes |K_L\rangle_r - |K_L\rangle_l \otimes |K_S\rangle_r
\big\rbrace \; .
\end{equation}
Then we get the state at time $t$ from (\ref{entangledKS}) by applying the
unitary operator
\begin{eqnarray}\label{U(t)unitary}
U(t,0) &=& U_l(t,0) \cdot U_r(t,0) \, ,
\end{eqnarray}
where the operators $U_l(t,0)$ and $U_r(t,0)$ act on the space of the left
and of the right mesons according to the time evolution (\ref{timeevolution}).
\\

What we are finally interested in are the quantum mechanical probabilities for
detecting, or not detecting, a specific quasi spin state on the left side
$|k_n\rangle_l$ and on the right side $|k_n\rangle_r$ of the source. For that
we need the projection operators $P_{l,r}(k_n)$ on the left, right quasi spin
states $|k_n\rangle_{l,r}$ together with the projection operators that act onto the
orthogonal states $Q_{l,r}(k_n)$
\begin{eqnarray}
P_l(k_n) \, &=& \, |k_n\rangle_{l\,l} \langle k_n| \qquad and \qquad
P_r(k_n) \;  =  \; |k_n\rangle_{r\,r} \langle k_n|\\
Q_l(k_n) \, &=& \, \mathbf{1} - P_l(k_n) \;\quad and \qquad
Q_r(k_n) \;  =  \; \mathbf{1} - P_r(k_n) \, .
\end{eqnarray}
So starting from the initial state (\ref{entangledKS}) the unitary
time evolution (\ref{U(t)unitary}) gives the state at a time $t_r$
\begin{eqnarray}
|\psi(t_r)\rangle &=& U(t_r,0)|\psi(t=0)\rangle \; = \; U_l(t_r,0) U_r(t_r,0)
|\psi(t=0)\rangle \, .
\end{eqnarray}
If we now measure a $k_m$ at $t_r$ on the right side means that we project onto
the state
\begin{eqnarray}
|\tilde{\psi}(t_r)\rangle &=& P_r(k_m) |\psi(t_r)\rangle \, .
\end{eqnarray}
This state, which is now a one-particle state of the left-moving particle, evolves until
$t_l$ when we measure a $k_n$ on the left side and we get
\begin{eqnarray}\label{evolutionexact}
|\tilde{\psi}(t_l, t_r)\rangle &=& P_l(k_n) U_l(t_l,t_r)
P_r(k_m) |\psi(t_r)\rangle \, .
\end{eqnarray}
The probability of the joint measurement is given by the squared
norm of the state (\ref{evolutionexact}). It coincides (due to unitarity,
composition laws and commutation properties of $l,r$-operators) with the
state
\begin{eqnarray}\label{evolutionfactorized}
|\psi(t_l,t_r)\rangle &=& P_l(k_n) P_r(k_m) U_l(t_l,0) U_r(t_r,0)
|\psi(t=0)\rangle \, ,
\end{eqnarray}
which corresponds to a factorization of the time into an eigentime $t_l$
on the left side and into an eigentime $t_r$ on the right side.

Then we can calculate the quantum mechanical probability
$P_{n,m}(Y, t_l; Y, t_r)$ for finding a $k_n$ at $t_l$ on the left side
{\it and} a $k_m$ at $t_r$ on the right side and the probability
$P_{n,m}(N, t_l; N, t_l)$ for finding $no$ such kaons by the following norms;
and similarly the probability $P_{n,m}(Y, t_l; N, t_r)$ when a $k_n$
at $t_l$ is detected on the left but $no\, k_m$ at $t_r$ on the right
\begin{eqnarray}
P_{n,m}(Y, t_l; Y, t_r) &=& ||P_l(k_n) P_r(k_m) U_l(t_l,0) U_r(t_r,0)
|\psi(t=0)\rangle||^2 \\
P_{n,m}(N, t_l; N, t_r) &=& ||Q_l(k_n) Q_r(k_m) U_l(t_l,0) U_r(t_r,0)
|\psi(t=0)\rangle||^2 \\
P_{n,m}(Y, t_l; N, t_r) &=& ||P_l(k_n) Q_r(k_m) U_l(t_l,0) U_r(t_r,0)
|\psi(t=0)\rangle||^2 \, .
\end{eqnarray}

\section{Bell inequalities for spin $1/2$ particles}\label{Section3}

In this section we will review briefly the well-known derivation of Bell-inequalities
\cite{bell2}. Our intention is to draw the readers attention to the analogies, but
more important to the differences of the spin/photon correlations as compared to the
quasi spin correlations discussed in the following sections.

We want to start with the derivation a general Bell inequality, the CHSH inequality,
named after Clauser, Horne, Shimony and Holt \cite{CHSH}, and then we derive from that
inequality - with two further assumptions - the original Bell inequality and the
Wigner-type inequality.

Let $A(n,\lambda)$ and $B(m,\lambda)$ be the definite values of two quantum
observables $A^{QM}(n)$ and $B^{QM}(m)$,
$\lambda$ denoting the hidden variables which are not accessible to an experimenter
but carry the additional information needed in a LRT.
The measurement result of one observable is $A(n,\lambda) = \pm 1$ corresponding to
the spin measurement 'spin up' and 'spin down' along the quantisation direction $n$
of particle $1$; and $A(n,\lambda) = 0$ if no particle was detected at all.
The analogue holds for the result $B(m,\lambda)$ of particle $2$.

Assuming now Bell's locality hypothesis ($A(n,\lambda)$ depends only on the direction
$n$, but not on $m$, the analogue holds for $B(m,\lambda)$)
-- which is the crucial point -- we have for the combined spin measurement the
following expectation value
\begin{eqnarray}\label{averagevalues}
M(n,m)=\int d\lambda\; \;\rho(\lambda) A(n,\lambda) B(m,\lambda)
\end{eqnarray}
with the normalized probability distribution
\begin{eqnarray}\label{norm}
\int d\lambda\; \;\rho(\lambda)=1\, .
\end{eqnarray}
This quantity $M(n,m)$ correspond to the quantum mechanical mean value
$M^{QM}(n, m)=\langle A^{QM}(n)\cdot B^{QM}(m)\rangle$.\\

A straight forward calculation (for example \cite{bell3,CHSH,Clauser}) gives the
estimate of the absolut value of the difference of two mean values
\begin{eqnarray}
| M(n,m)-M(n,m')| &\leq& \int d\lambda \;\rho(\lambda)\;\big\lbrace 1\pm A(n',\lambda)
B(m',\lambda)\big\rbrace\nonumber\\
& &\qquad+\int d\lambda \;\rho(\lambda)\,\big\lbrace 1\pm A(n',\lambda)
B(m,\lambda)\big\rbrace
\end{eqnarray}
and using normalization (\ref{norm}) we get
\begin{eqnarray}\label{chsh-inequality-derivation}
| M(n,m)-M(n,m')|\;\leq 2\;\pm| M(n',m')+M(n',m)|
\end{eqnarray}
and more symmetrically
\begin{eqnarray}\label{chsh-inequality}
| M(n,m)-M(n,m')|+| M(n',m')+M(n',m)|\; \leq \;2 \;.
\end{eqnarray}
This is the familiar CHSH-inequality, derived by Clauser, Horne, Shimony and Holt \cite{CHSH}
in 1969. Every local realistic hidden variable theory must obey that inequality.\\
\\
Inserting the quantum mechanical expectation values $M^{QM}(n,m)$ for $M(n,m)$, we
get with $\phi_{n,m}$ being the angle between the two quantization directions $n$ and $m$
\begin{eqnarray}\label{chshphoton}
S(n, m, n', m')=|\cos(\phi_{n, m})-\cos(\phi_{n,
m'})|+|\cos(\phi_{n',m'})+\cos(\phi_{n',m})|\;\leq\;2 \, ,
\end{eqnarray}
which is for some choices of the angles $\phi$ violated; the maximal value of the left
hand side is $2 \sqrt{2}$, with for instance $\phi_{n,m'}=\frac{3 \pi}{4}$ and
$\phi_{n,m}=\phi_{n',m'}=\phi_{n',m}=\frac{\pi}{4}$.
Experimentally, for entangled photon pairs inequality (\ref{chshphoton}) is violated
under strict Einstein locality conditions in an impressive way, with a result
close in agreement with QM \cite{zeilinger}, confirming such previous experimental
results on similar inequalities \cite{FreedmanClauser,FryThompson,aspect}.\\

In order to come to the original Bell inequality or to the Wigner inequality we make
two assumptions, first we assume always perfect correlation $M(n,n)=-1$ and second
the measurement of the state of the particles has to be perfect, so there are no
omitted events which were interpreted in the CHSH derivation as $0$ results.

Considering now just $3$ different quantization directions, so choosing e.g. $n'=m'$,
inequality (\ref{chsh-inequality-derivation}) gives
\begin{eqnarray}\label{Bell-orginal}
& &| M(n,m)-M(n,n')|\;\leq 2\;\pm\{ \underbrace{M(n',n')}+M(n',m)\}\nonumber\\
& &\hphantom{| M(n,m)-M(n,n')|\;\leq 2\;\pm\{ }\;\;-1\;\forall\; n'\nonumber\\
& &\textrm{or}\nonumber\\
& &| M(n,m)-M(n,n')|\;\leq\; 1+M(n',m).
\end{eqnarray}
This is the famous original inequality derived by J.S. Bell \cite{bell} in 1964.
Note, that this derivation is already true for the entangled kaon system
where the different kaon quasi spin eigenstates on the left and right side, measured
at equal times, play the role of the different angles,
see Section \ref{Wigner-typeSection}.\\

Finally, we rewrite the expectation value for two spin $1/2$ particles in terms of
probabilities
\begin{eqnarray}
M(n,m)
&=& P(\vec{n} \Uparrow; \vec{m} \Uparrow)+P(\vec{n} \Downarrow; \vec{m} \Downarrow)
-P(\vec{n} \Uparrow; \vec{m} \Downarrow)-P(\vec{n} \Downarrow; \vec{m}
\Uparrow)\nonumber\\
&=& -1+4\; P(\vec{n} \Uparrow; \vec{m} \Uparrow)\; ,
\end{eqnarray}
where we used $\sum P=1$.
Then Bell's original inequality (\ref{Bell-orginal}) provides the Wigner inequality
\begin{eqnarray}
P(\vec{n}; \vec{m})\;\leq\;P(\vec{n}; \vec{n'})+P(\vec{n'}; \vec{m})\; ,
\end{eqnarray}
where the $P$ can be the measurement of all spins up or down on both sides,
or spin up on one side and spin down on the other side, or vice versa.
Note, that the Wigner inequality has been originally derived by a set-theoretical
approach.

\section{Generalized Bell inequalities for $K$-mesons}

Let us consider again the entangled state $| \psi (t=0) \rangle$
(\ref{entangledK0}) of a \kkk pair and its time evolution $U(t,0) | \psi (0) \rangle$,
then we find the following situation:

Performing two measurements to detect the kaons at the same time
at the left side and at the right side of the source the probability
of finding two mesons with the same strangeness $K^0 K^0$ or
$\bar K^0 \bar K^0$ is zero.
If we measure at time $t$ a $\bar K^0$ meson on the left side, we will
find with certainty at the same time $t$ $no \, \bar K^0$ on the right side.
This is an EPR-Bell correlation analogously to the spin $1/2$ or photon
(e.g., with polarization vertical - horizontal) case.
The analogy would be perfect, if the kaons were stable
($\Gamma _S = \Gamma _L = 0$); then the quantum probabilities become
\begin{eqnarray}
P(Y,t_l;Y,t_r) &=& P(N,t_l;N,t_r) \; = \; \frac{1}{4}
\big\lbrace 1 - \cos(\Delta m(t_l - t_r))\big\rbrace \nonumber\\
P(Y,t_l;N,t_r) &=& P(N,t_l;Y,t_r) \; = \; \frac{1}{4}
\big\lbrace 1 + \cos(\Delta m(t_l - t_r))\big\rbrace \, .
\end{eqnarray}
They coincide with the probabilities of finding simultaneously two
entangled spin $1/2$ particles in spin directions $\Uparrow$ $\Uparrow$
or $\Uparrow$ $\Downarrow$ along two chosen directions $\vec n$
and $\vec m$
\begin{eqnarray}
P(\vec n,\Uparrow ;\vec m,\Uparrow) &=& P(\vec n,\Downarrow ;\vec m,\Downarrow)
\; = \; \frac{1}{4}
\big\lbrace 1 - \cos \theta \big\rbrace \nonumber\\
P(\vec n,\Uparrow ;\vec m,\Downarrow) &=& P(\vec n,\Downarrow ;\vec m,\Uparrow)
\; = \; \frac{1}{4}
\big\lbrace 1 + \cos \theta \big\rbrace \, .
\end{eqnarray}
The time differences $\Delta m(t_l - t_r)$ in the kaon case play
the role of the angle differences $\theta$ in the spin $1/2$ case.

Nevertheless, there are important physical differences between
kaon and spin $1/2$ states (for an experimenter's point of view, see Ref.\cite{Gisin}).
\begin{enumerate}
\item While in the spin $1/2$ or photon case one can test whether
a system is in an arbitrary spin state
$\alpha |\Uparrow\rangle + \beta |\Downarrow\rangle$
one cannot test it for an arbitrary superposition
$\alpha |K^0\rangle + \beta |\bar K^0\rangle$.

\item For entangled spin $1/2$ particles or photons it is sufficient
to consider the direct product space $H^l_{spin} \otimes H^r_{spin} \,$,
however, this is not so for kaons. The unitary time evolution of a
kaon state also involves the decay product states (see Chapter 3),
therefore one has to include the decay product spaces which are
orthogonal to the product space $H^r_{kaon} \otimes H^l_{kaon} \,$.
\end{enumerate}

So by measuring a $\bar K^0$ at the left side we can predict
with certainty to find at the same time $no \, \bar K^0$ at the right
side. In any LRT this property $no \, \bar K^0$ must be present at the
right side independent of having the measurement performed or not.
In order to discriminate between QM and LRT we set up a Bell inequality
for the kaon system where now the different times play the role of the
different angles in the spin $1/2$ case.
But, in addition, we use the freedom of choosing a particular quasi
spin state of the kaon, the strangeness eigenstate, the mass eigenstate,
or the $CP$ eigenstate.

\subsection{Expectation values and locality}

As discussed before in kaon systems we have the freedom of  choosing the time, when
a measurement takes place {\it and} the freedom which
particular quasi spin state we want to measure.

The locality hypothesis then requires that the results of measurement
on the left side are completely independent of the chosen time
and chosen quasi spin state in the measurement on the right side.

Let us consider an observable $O(k_n, t_a)$ on each side of
the source, which gets the value $+1$ if in a measurement at
time $t_a$ the quasi spin state $k_n$ is found and the value $-1$
if not.

Then we can define a correlation function $O(k_n, t_a;k_m, t_b)$
which gets the value $+1$, both when at the left side a $k_n$
at $t_a$ and at the right side a $k_m$ at $t_b$ was detected or
when $no\, k_n$ and $no\, k_m$ was found. In the case that
only one of the desired quasi spin eigenstates has been found,
no matter at which side, the correlation function has the value
$-1$.\\

{\bf Locality hypothesis:}
Locality in the sense of Bell means that the correlation function
$O(k_n,t_a;k_m,t_b)$ is equal to the product of the observables
on each side

\begin{equation}
O(k_n, t_a; k_m, t_b) \, =\, O^l(k_n, t_a)\;\cdot\; O^r(k_m, t_b)\, .
\end{equation}
\\
Then the following relation holds
\begin{eqnarray}\label{observablerelation}
|O(k_n, t_a; k_m, t_b)- O(k_n, t_a; k_{m'}, t_d)|\; +
\;|O(k_{n'}, t_c; k_{m'}, t_d)+O(k_{n'}, t_c; k_m, t_b)|\; = \; 2,\nonumber\\
\end{eqnarray}
with $k_n, k_m, k_{n'}$ and $k_{m'}$ being arbitrary quasi spin eigenstates of the meson
and $t_a, t_b, t_c$ and $t_d$ four different times.

Now we consider a series of $N$ identical measurements and we denote by $O_i$
the value of $O$ in the $i$-th experiment. The average is given by
\begin{eqnarray}\label{meanvalue}
M(k_n, t_a; k_m, t_b) &=& \frac{1}{N} \sum_{i=1}^N O_i(k_n, t_a; k_m, t_b) \, .
\end{eqnarray}
Taking the absolute values of differences and sums of such
averages and inserting relation (\ref{observablerelation}) we obtain the Bell-CHSH
inequality for the expectation values
\begin{equation}\label{chshmeanvalue}
|M(k_n, t_a; k_m, t_b) - M(k_n, t_a; k_{m'}, t_d)| +
| M(k_{n'}, t_c; k_{m'}, t_d) + M(k_{n'}, t_c; k_m ,t_b)|\; \leq \; 2 \, .
\end{equation}
If we identify $M(k_n, t_a; k_m, t_b)\equiv M(n,m)$ we are back at
the inequality (\ref{chsh-inequality}) for the spin $1/2$ case.

\subsection{Probabilities}

Now we consider the expectation value (\ref{meanvalue}) for the series of
identical measurements in terms of the probabilities, where we
denote by $P_{n,m}(Y, t_a; Y, t_b)$ the probability for finding a
$k_n$ at $t_a$ on the left side and finding a $k_m$ at $t_b$ on the
right side and by $P_{n,m}(N, t_a; N, t_b)$ the probability for finding
$no$ such kaons; similarly $P_{n,m}(Y, t_a; N, t_b)$ denotes the
case when a $k_n$ at $t_a$ is detected on the left but $no\, k_m$ at
$t_b$ on the right. Then we can re-express the expectation value
by the following linearcombination
\begin{eqnarray}\label{meanvalueprob}
M(k_n, t_a; k_m, t_b)&=& P_{n,m}(Y, t_a; Y, t_b) + P_{n,m}(N, t_a; N, t_b)\nonumber\\
& & - P_{n,m}(Y, t_a; N, t_b) - P_{n,m}(N, t_a; Y, t_b) \, .
\end{eqnarray}
Since the sum of the probabilities for $(Y,Y)$, $(N,N)$, $(Y,N)$ and
$(N,Y)$ must be unity we get
\begin{eqnarray}
M(k_n, t_a; k_m, t_b) &=& -1 + 2\, \big\lbrace P_{n,m}(Y, t_a; Y, t_b)
+ P_{n,m}(N, t_a; N, t_b)\big\rbrace \, .
\end{eqnarray}
Note that relation (\ref{meanvalueprob}) between the expectation value
and the probabilities is satisfied for QM and LRT as well.

Setting this expression into the Bell-CHSH inequality (\ref{chshmeanvalue})
we finally arrive at the following inequality for the probabilities
\begin{eqnarray}\label{CHSHprobab}
& &|P_{n,m}(Y, t_a; Y, t_b) + P_{n,m}(N, t_a; N, t_b) -
P_{n,m'}(Y, t_a; Y, t_d) - P_{n,m'}(N, t_a; N, t_d)|\; \leq \; \nonumber\\
& &\hphantom{\qquad \qquad}
1 \pm \big\{ -1 + P_{n',m}(Y, t_c; Y, t_b) + P_{n',m}(N, t_c; N, t_b)\nonumber\\
& &\hphantom{\qquad \qquad \qquad \quad}
+ P_{n',m'}(Y, t_c; Y, t_d) +  P_{n',m'}(N, t_c; N, t_d)\big\}
\end{eqnarray}
or
\begin{eqnarray}\label{sfunction}
\lefteqn{S(k_n, k_m, k_{n'}, k_{m'}; t_a, t_b, t_c, t_d)=}\nonumber\\
& &|P_{n,m}(Y, t_a; Y, t_b) + P_{n,m}(N, t_a; N, t_b) -
P_{n,m'}(Y, t_a; Y, t_d) - P_{n,m'}(N, t_a; N, t_d)| +\nonumber\\
& &|-1 + P_{n',m}(Y, t_c; Y, t_b) + P_{n',m}(N, t_c; N, t_b)\nonumber\\
& &\hphantom{\qquad }
+ P_{n',m'}(Y, t_c; Y, t_d) + P_{n',m'}(N, t_c; N, t_d)|\;\leq\; 1\, .
\end{eqnarray}

\subsection{Wigner-type inequalities}\label{Wigner-typeSection}

What we aim is to find Wigner-type inequalities. The most general one
we get from above inequality (\ref{CHSHprobab}) by choosing the upper sign $+$
\begin{eqnarray}\label{wignergeneral}
P_{n, m}(Y, t_a; Y, t_b) &\;\leq\;& P_{n,m'}(Y, t_a; Y, t_d) +
P_{m',n'}(Y, t_d; Y, t_c) + P_{n',m}(Y, t_c ; Y, t_b)\nonumber\\
& & + \, h(n,m,n',m';t_a,t_b,t_c,t_d)
\end{eqnarray}
where
\begin{eqnarray}\label{correctionfunction}
h(n,m,n',m';t_a,t_b,t_c,t_d) &=& - P_{n,m}(N, t_a; N, t_b) +
P_{n,m'}(N, t_a; N, t_d)\nonumber\\
& & + P_{n',m}(N, t_c; N, t_b) + P_{n',m'}(N, t_c; N, t_d)
\end{eqnarray}
is a correction function to the usual set-theoretical result, see Section \ref{Section3}. It arises
because for a unitary time evolution we also have to include the
decay states (see Eq.(\ref{timeevolution})), contributing to the $no\, kaon$ states,
thus the decay product spaces which are orthogonal to the product
space $H^r_{kaon} \otimes H^l_{kaon}$\, .\\

For zero times $t_{a,b} \to 0$, when we have no decays, the probabilities
for $(N,N)$ become the ones for $(Y,Y)$
\begin{equation}
P_{n,m}(N, t_a; N, t_b)|_{t_{a,b}=0} \equiv P_{n,m}(Y, t_a; Y, t_b)|_{t_{a,b}=0}\, ,
\end{equation}
the correction function (for $t_a=t_b=t_c=t_d=t=0$) is then equal to
\begin{eqnarray}
h(n,m,n',m';t=0) &=& - P_{n,m}(Y,Y)|_{t=0} + P_{n,m'}(Y,Y)|_{t=0} +
P_{n',m}(Y,Y)|_{t=0}\nonumber\\
& & + P_{n',m'}(Y,Y)|_{t=0}\, ,
\end{eqnarray}
and just adds up to the inequality (\ref{wignergeneral}) in such a way that we
obtain the usual set-theoretical result
\begin{eqnarray}\label{wignergeneral-t0}
P_{n, m}(Y,Y)|_{t=0} &\;\leq\;& P_{n,m'}(Y,Y)|_{t=0} +
P_{m',n'}(Y,Y)|_{t=0} + P_{n',m}(Y,Y,)|_{t=0}\, .
\end{eqnarray}

Of course, the case we are interested in contains only 3 different
states, so we put $n'=m'$ and $t_c=t_d\, ,$ then the probability for $(Y,Y)$
vanishes $P_{n',n'}(Y, t_c; Y, t_c) =~0$  due to the EPR-Bell anticorrelation
but certainly not the probability for $(N,N)$ $P_{n',n'}(N, t_c; N, t_c) \not= 0$
(it vanishes only for $t_c \to 0$).

So we obtain the following Wigner-type inequality for 3 different
quasi spin states
\begin{eqnarray}\label{wignerspecial}
P_{n, m}(Y, t_a; Y, t_b) &\;\leq\;& P_{n,n'}(Y, t_a; Y, t_c) +
P_{n',m}(Y, t_c ; Y, t_b)\nonumber\\
& & + \, h(n,m,n';t_a,t_b,t_c)
\end{eqnarray}
with the correction function
\begin{eqnarray}\label{correctionfunctionspecial}
h(n,m,n';t_a,t_b,t_c) &=& - P_{n,m}(N, t_a; N, t_b) + P_{n,n'}(N, t_a; N, t_c)
\nonumber\\
& & + P_{n',m}(N, t_c; N, t_b) + P_{n',n'}(N, t_c;N, t_c).
\end{eqnarray}
Again, in the limit of zero times $t \to 0$ we arrive at the familiar
Wigner-type inequality
\begin{equation}
P_{n,m}(Y,Y)|_{t=0}\;\leq\;P_{n,n'}(Y,Y)|_{t=0} + P_{n',m}(Y,Y)|_{t=0}\, .
\end{equation}

We certainly also can achieve Bell's original case (\ref{Bell-orginal}),
which is more restrictive since we have to require perfect
anticorrelation
\begin{equation}
M(k_n, t; k_n, t) \; = \; -1 \, .
\end{equation}
Then the general CHSH relation, Eq.(\ref{chshmeanvalue}), implies the
specific inequality of Bell
\begin{equation}\label{bellinequal}
|M(k_n, t; k_m, t) - M(k_n, t; k_{n'}, t)| \; \leq \; 1 + M(k_{n'}, t; k_m ,t) \, .
\end{equation}
Converting it into a Wigner-type we come back to inequality
(\ref{wignerspecial}), but with a smaller correction function
\begin{equation}
h_{Bell}(t) \; = \; h_{CHSH}(t) - P_{n',n'}(N, t; N, t) \, ,
\end{equation}

which is more restrictive.

\subsection{The choice sensitive to the $CP$ parameter $\varepsilon$}

Choosing the quasi spin states
\begin{eqnarray}\label{Uchoice}
& &|k_n\rangle\; = |K_S\rangle\nonumber\\
& &|k_m\rangle = |\bar K^0\rangle\nonumber\\
& &|k_{n'}\rangle = |K_1^0\rangle\, ,
\end{eqnarray}
and denoting the probabilities
$P_{K_S,\bar K^0}(Y,Y)|_{t=0}\equiv P(K_S,\bar K^0)$ etc.,
we recover Uchiyama's inequality \cite{Uchiyama}
\begin{equation}
P(K_S,\bar K^0)\; \leq\; P(K_S,K_1^0) + P(K_1^0,\bar K^0)
\end{equation}
which he derived by a set-theoretical approach. The interesting
point here is its connection to a physical parameter, the
$CP$ violating parameter $\varepsilon$. As Uchiyama has shown
his inequality can be turned into an inequality for $\varepsilon$
\begin{equation}\label{Wignerepsilon}
Re\{\varepsilon\}\;\leq\;|\varepsilon|^2
\end{equation}
which is obviously violated by the experimental value of
$\varepsilon$, having an absolute value of about $10^{-3}$ and
a phase of about $45^\circ$ \cite{ParticleData}.

An other meaningful choice would be the replacement of the short
living state $|K_S\rangle$ by the long living state $|K_L\rangle$
and the $CP$ eigenstate $|K_1\rangle$ by $|K_2\rangle$ in Eq.(\ref{Uchoice})
then we arrive at the same inequality (\ref{Wignerepsilon}).\\

Our Wigner-type inequality (\ref{wignerspecial}) differs from the
ones discussed in the literature
\cite{domenico,Benatti98,Benattiepsilonstrich,Bramon,AncoBramon,Gisin}; in the sense that we have an additional term $h$
(\ref{correctionfunctionspecial}) due to the unitary time
evolution of the considered states. Since $h$
is positive it worsens the possibility for quantum mechanics to
violate the Bell inequality.

This can be clearly seen in case of equal times $t_a=t_b=t_c=t$,
when the exponential $t$-dependence factorizes in the $(Y,Y)$
probabilities but not in the $(N,N)$ ones. Then we have for the
choice (\ref{Uchoice}) the following Wigner-type inequality
\begin{equation}\label{wignerspecialequaltime}
e^{-2\Gamma t} P(K_S,\bar K^0)\; \leq\; e^{-2\Gamma t} P(K_S,K_1^0)
+ e^{-2\Gamma t} P(K_1^0,\bar K^0) + h(K_S,\bar K^0,K_1^0;t)\, ,
\end{equation}
where the probabilities and the correction function $h$ can be
found explicitly in the Appendix \ref{appendixuchy}. As we see due to the fast
damping of the probabilities (and $h \to 2$) a violation of
inequality (\ref{wignerspecialequaltime}) by QM is only possible for
very small times, in fact, only for times $t\leq 8\cdot10^{-4}\, \tau_S\,$.

But fortunately there exist certain cases where the situation is
better. We can avoid a fast increase of the correction function $h$
by taking the times $t_a=t_c$ and $t_a\leq t_b$. Then a violation
of the Wigner-type inequality (\ref{wignerspecial}) occurs, which is
strongest for $t_a\approx 0$; and in this case $t_b$ can be chosen up
to $t_b\leq 4\, \tau_S\,$, which is already quite large.\\

\subsection{The choice sensitive to the direct $CP$ parameter $\varepsilon'$}

As shown by Benatti and Floreanini in Refs.\cite{Benatti98,Benattiepsilonstrich}, the
case has been also discussed carefully in Refs.\cite{Bramon,AncoBramon}, some
decay end-products can be identified with the quasi spin eigenstates. For example, the
two neutral pions or the two charged pions can be associated with the quasi spin
eigenstates:
\begin{eqnarray}
&|K_{00}\rangle\;=\frac{1}{\sqrt{1+|\varepsilon_{00}|}}\{ |K_1^0\rangle + \varepsilon_{00}
|K_2^0\rangle\}&\;\longrightarrow\;|\pi^0 \pi^0\rangle\nonumber\\
&|K_{+-}\rangle=\frac{1}{\sqrt{1+|\varepsilon_{+-}|}}\{ |K_1^0\rangle + \varepsilon_{+-}
|K_2^0\rangle\}&\;\longrightarrow\;|\pi^+ \pi^-\rangle
\end{eqnarray}
with
\begin{eqnarray}\label{epsilon00}
\varepsilon_{00}\;&=&-2\varepsilon'+i \frac{Im\{\mathcal{A}_0\}}{Re\{\mathcal{A}_0\}}\nonumber\\
\varepsilon_{+-}&=&\varepsilon'+i \frac{Im\{\mathcal{A}_0\}}{Re\{\mathcal{A}_0\}}\;.
\end{eqnarray}

Here $\mathcal{A}_0\equiv\langle \pi\pi,I=0|H_w|K^0\rangle$ is the weak decay amplitude
with $I$ being the isospin (for further
information, see Refs.\cite{Nachtmann,HoKim,Branco}) and $\varepsilon'$ being the direct
$CP$ violation
parameter; the third order and higher orders in $\varepsilon$ and
$\varepsilon_{00}, \varepsilon_{+-}$ are already neglected.

We choose -- analogously to previous section -- the quasi spin states
\begin{eqnarray}\label{Bchoice}
& &|k_n\rangle=|K^0\rangle\nonumber\\
& &|k_m\rangle=|K_{00}\rangle\nonumber\\
& &|k_{n'}\rangle= |K_{+-}\rangle
\end{eqnarray}
and we get the following Wigner-type inequality for $t=0$
\begin{eqnarray}
P(K^0, K_{00})\;\leq\; P(K^0, K_{+-})+P(K_{+-}, K_{00})\;.
\end{eqnarray}

The  calculation of the probabilities gives an inequality
\begin{eqnarray}
|-Re\{\varepsilon_{00}\}(1+|\varepsilon_{+-}|^2)+Re\{\varepsilon_{+-}\}(1+|
\varepsilon_{00}|^2)|\;\leq |\varepsilon_{00}|^2+|\varepsilon_{+-}|^2-2
Re\{\varepsilon_{00}^*\varepsilon_{+-}\}\nonumber\\
\end{eqnarray}
which, when the results (\ref{epsilon00}) for $\varepsilon_{00}$ and $\varepsilon_{+-}$
are inserted,
turns into an inequality in the direct $CP$ violating parameter $\varepsilon'$
(third order terms neglected)
\begin{eqnarray}
Re\{\varepsilon'\}\;\leq\;3|\varepsilon'|^2\;,
\end{eqnarray}
the inequality of Refs.\cite{Benatti98,Benattiepsilonstrich}.

This inequality is clearly violated by the experimental value of $\varepsilon'$,
$|\varepsilon'|\lesssim 10^{-6}$  and has a phase of about $45^\circ$ \cite{ParticleData}.

Again, for times $t>0$ we have to include the correction function h. Choosing all
four times equal $t_a=t_b=t_c=t_d=t$, the inequality (\ref{wignerspecial}) with the choice
(\ref{Bchoice}) cannot be violated for times
larger than $t=3.7\cdot 10^{-6} \tau_S$.

Varying all four times, unfortunately, does not improve the test QM versus LRT, we only find a
violation in the region where all times are smaller than $10^{-6}\tau_S$.

\subsection{The choice of the strangeness eigenstate}\label{ghirardisection}

Finally we also can reproduce the case of Ghirardi, Grassi and Weber
\cite{ghirardi91}, we just have to consider the
same quasi spin states
\begin{eqnarray}
k_n=k_m=k_{n'}=k_{m'}= \bar K^0.
\end{eqnarray}
Evaluating the Bell-CHSH inequality (\ref{chshmeanvalue}) by the
quantum mechanical probabilities, neglecting $CP$ violation, the
result is \cite{ghirardi91}
\begin{eqnarray}\label{chshghirardietal}
& &|e^{-\frac{\Gamma_S}{2} (t_a+t_c)} \, \cos(\Delta m (t_a-t_c))
- e^{-\frac{\Gamma_S}{2} (t_a+t_d)} \, \cos(\Delta m (t_a-t_d))|\nonumber\\
& & + |e^{-\frac{\Gamma_S}{2} (t_b+t_c)} \, \cos(\Delta m (t_b-t_c))
+ e^{-\frac{\Gamma_S}{2} (t_b+t_d)} \, \cos(\Delta m (t_b-t_d))|\;\leq\;2\, .
\end{eqnarray}
Unfortunately, inequality (\ref{chshghirardietal}) \emph{cannot}
be violated \cite{ghirardi91,ghirardi92} for any choice of the four (positive)
times $t_a, t_b, t_c, t_d$ due to the interplay between the kaon decay and
strangeness oscillations. As demonstrated in Ref. \cite{trixi}
a possible violation depends very much on the kaon parameter
$x = \Delta m /\Gamma\;$; if we had $x=4.3$ instead of the
experimental $x\approx 1$, this Bell-CHSH inequality
(\ref{chshghirardietal}) would be broken. Note, that in this case the CHSH inequality
maximizes at different time values than expected from the corresponding photon CHSH
inequality (\ref{chshphoton}).

\section{Summary and Conclusions}

\begin{itemize}
\item \textbf{Quantum theory}\\
We consider the time evolution of neutral kaons and emphasize the unitary time
evolution which includes the decay states. Starting at $t=0$ with a $K^0$ one gets after
a certain time $t$ a superposition of the strangeness eigenstates due to strangeness
oscillations \emph{and} the decay
states. In this way we consider the total Hilbert space -- analogously to the photon
case.

Then we treat entangled states and derive their quantum mechanical probabilities
of finding or not finding arbitrary quasi spin states at arbitrary times. With these
QM probabilities we calculate the quantum mechanical expectation value.

\item \textbf{LRT}\\
We derive the general Bell-CHSH inequality (\ref{chshmeanvalue}) based on a local realistic hidden
variable theory. From this general Bell inequality follows a Wigner-type inequality
(\ref{wignergeneral}) and an inequality analogously to Bell's original version.

\item \textbf{QM versus LRT}\\
Next we compare the quantum theory with LRT that means we insert the quantum
mechanical expectation value into the general Bell-CHSH inequality. Expressing the
expectation value in terms of probabilities we arrive at a Wigner-type inequality
(\ref{wignergeneral}) which contains an additional term due to the unitary time evolution, the
correction function $h$ (\ref{correctionfunction}).This function $h$ is missing in the inequalities of
other authors \cite{domenico,Benatti98,Benattiepsilonstrich} since they restrict themselves
to a sub-set of the Hilbert space.

\item \textbf{Results}\\
This correction function $h$ makes it rather difficult for QM to violate the Bell
inequality (in order to show the non-local character of QM). In case of Ghirardi,
Grassi and Weber \cite{ghirardi91}, where only $\bar K^0$ or no $\bar K^0$ is detected, it is
impossible for any choice
of the times that QM violates the BI. On the other hand, if we consider in addition to
the choice of time the freedom of choosing particular quasi spin eigenstates, then we find
cases were QM does violate the BI for certain times. For example, in the choice
(\ref{Uchoice}) the Bell inequality is violated for $t_a=t_c\approx 0$ and $t_b\leq
4\tau_S$. Considering another choice (\ref{Bchoice}) we find no violation at all, except for
$t=0$.

\item \textbf{Comments}\\
The authors of Refs.\cite{domenico,Benatti98,Benattiepsilonstrich,Gisin} restrict their analysis to a
subset of the Hilbert space; tests on such subspaces, however, probe only a
restricted class of LRT. In such subspaces Bell inequalities may be violated, but
this need not to be the case in the total Hilbert space.

We, on the other hand, aim to exclude the largest class of LRT, therefore we work
with a unitary time evolution, a point of view we share with \cite{ghirardi91,ghirardi92}.

\item \textbf{Outlook}\\
In cases, where QM does not violate the Bell inequality, we trace it back to the
specific value of the internal parameter $x=\frac{\Delta m}{\Gamma}$, given by
Nature. And it does not indicate that these massive systems \emph{have} real properties
independent of the act of measurement. However, some of these quasi spin eigenstates
are difficult to detect experimentally, in this connection the idea of the ``quasi spin
rotations'', introducing appropriate kaon ``regenerators'' along the kaon flight
paths, and the resulting Bell inequalities is of special interest
(see e.g. \cite{Bramon,AncoBramon}).

An interesting feature of the neutral kaon systems in comparison with photon is
that this system has $CP$ violation. Although the Bell inequalities themselves are hard to
check experimentally, they imply an inequality on the physical $CP$ violation parameter
$\varepsilon$ or $\varepsilon'$, which is experimentally testable.
\end{itemize}

\section{Acknowledgement}

The authors want to thank W. Grimus, N. Gisin and G.C. Ghirardi for fruitful discussions
and suggestions. One of the authors, B.C. Hiesmayr, was supported by Austrian FWF project
P14143-PHY and the Austrian-Czech Republic Scientific Collaboration, project KONTAKT 1999-8.

\newpage
\appendix
\section{}\label{Qmprobabilities}

\subsection{Formula for the choice sensitive to the $CP$ parameter
$\varepsilon$}\label{appendixuchy}

The $(Y,Y)$-probabilities:
\begin{eqnarray}
\lefteqn{P_{K_S, \bar K^0}(Y, t_l; Y, t_r)=}\nonumber\\
& &N_{SL}^2 \frac{1}{4} (1-\delta)
\biggl\lbrace e^{-\Gamma_S t_l-\Gamma_L t_r}+ \delta^2 e^{-\Gamma_L t_l-\Gamma_S t_r}+
2 \delta \cos(\Delta m \Delta t) \cdot e^{-\Gamma (t_l+t_r)}
\biggr\rbrace\nonumber\\
& &\\
\lefteqn{P_{K_S,K_1^0}(Y, t_l; Y, t_r)=}\nonumber\\
& &N_{SL}^2\frac{1}{2}\frac{1}{1+|\varepsilon|^2}\biggl\lbrace
|\varepsilon|^2 e^{-\Gamma_S t_l-\Gamma_L t_r}+
\delta^2 e^{-\Gamma_S t_r-\Gamma_L t_l}-
2\delta Re\{\varepsilon^* e^{-i \Delta m \Delta t}\}\cdot e^{\Gamma (t_l+t_r)}
\biggr\rbrace\nonumber\\
& &\\
\lefteqn{P_{K_1^0, \bar K^0}(Y, t_l; Y, t_r)=}\nonumber\\
& &N_{SL}^2\frac{1}{4}\frac{1-\delta}{1+|\varepsilon|^2}
\biggl\lbrace
e^{-\Gamma_S t_l-\Gamma_L t_r}+|\varepsilon|^2 e^{-\Gamma_L t_l-\Gamma_S t_r}+ 2
Re\{\varepsilon e^{-i\Delta m \Delta t}\}\cdot e^{-\Gamma(t_l+t_r)}
\biggr\rbrace\nonumber\\
& &\\
\lefteqn{P_{K_1^0, K_1^0}(Y, t_l; Y, t_r)=}\nonumber\\
& &N_{SL}^2\frac{1}{2}
\frac{|\varepsilon|^2}{(1+|\varepsilon|^2)^2}\biggl\lbrace
e^{-\Gamma_S t_l-\Gamma_L t_r}+e^{-\Gamma_L t_l-\Gamma_S t_r}-2 \cos(\Delta m \Delta
t) \cdot e^{-\Gamma (t_l+t_r)}\biggr\rbrace
\end{eqnarray}

The correction function:

\begin{eqnarray}
\lefteqn{h(K_S, t_a; \bar K^0, t_b; K_1^0, t_c; K_1^0, t_d)=}\nonumber\\
& &-P_{K_S, \bar K^0}(Y, t_a; Y,
t_b)+P_{K_S, K_1^0}(Y ,t_a; Y ,t_c)+P_{K_1^0,K_1^0}(Y ,t_d; Y ,t_b)+P_{K_1^0,\bar K^0}(Y ,t_d;
Y,
t_c)\nonumber\\
&+&3-N_{SL}^2\biggl\lbrace e^{-\Gamma_S t_a}+\delta^2 e^{-\Gamma_L
t_a}-2\delta^2 \cos(\Delta m t_a)\cdot e^{-\Gamma t_a}\nonumber\\
& &\hphantom{+3-N_{SL}^2}+
\frac{1}{1+|\varepsilon|^2}\big(e^{-\Gamma_S t_d}+|\varepsilon|^2 e^{-\Gamma_L
t_d}-2 \delta Re\{\varepsilon\; e^{-i \Delta m t_d}\}\cdot e^{-\Gamma t_d}\big)\nonumber\\
& &\hphantom{+3-N_{SL}^2}+
\frac{1-\delta}{2} \big( e^{-\Gamma_S t_b}+e^{-\Gamma_L
t_b}+2\delta\cos(\Delta m t_b)\cdot e^{-\Gamma
t_b}\big)\nonumber\\
& &\hphantom{+3-N_{SL}^2}+
\frac{1}{1+|\varepsilon|^2} \big(e^{-\Gamma_S t_c}+|\varepsilon|^2 e^{-\Gamma_L
t_c}-2\delta Re\{ \varepsilon\; e^{-i \Delta m t_c}\}\cdot e^{-\Gamma
t_c}\big)\biggr\rbrace
\end{eqnarray}

\newpage
\subsection{Formula for the choice sensitive to the direct $CP$ parameter $\varepsilon'$}

The $(Y,Y)$ probabilities:

\begin{eqnarray}
\lefteqn{P_{K^0,K_{00}}(Y, t_l; Y, t_r)=\frac{N_{SL}}{2} \frac{1+\delta}{2}\;
\frac{1}{1+|\varepsilon|^2}\;\frac{1}{1+|r_{00}|^2}}\nonumber\\
& &\biggl\lbrace|\varepsilon_{00}^*+\varepsilon|^2\; e^{-\Gamma_S t_l-\Gamma_L t_r}+
|1+\varepsilon\; \varepsilon_{00}^*|^2\; e^{-\Gamma_L t_l-\Gamma_S t_r}\nonumber\\
& &-2 Re\{(\varepsilon_{00}+\varepsilon^*)(1+\varepsilon\; \varepsilon_{00}^*)\;e^{-i
\Delta m \Delta t}\}\cdot e^{-\Gamma (t_l+t_r)}\biggr\rbrace\\
\lefteqn{P_{K^0,K_{+-}}(Y, t_l; Y, t_r)=\frac{N_{SL}}{2} \frac{1+\delta}{2}\;
\frac{1}{1+|\varepsilon|^2}\;\frac{1}{1+|r_{+-}|^2}}\nonumber\\
& &\biggl\lbrace|\varepsilon_{+-}^*+\varepsilon|^2\; e^{-\Gamma_S t_l-\Gamma_L t_r}+
|1+\varepsilon\; \varepsilon_{+-}^*|^2\; e^{-\Gamma_L t_l-\Gamma_S t_r}\nonumber\\
& &-2 Re\{(\varepsilon_{+-}+\varepsilon^*)(1+\varepsilon\; \varepsilon_{+-}^*)\;e^{-i
\Delta m \Delta t}\}\cdot e^{-\Gamma (t_l+t_r)}\biggr\rbrace\\
\lefteqn{P_{K_{+-},K_{00}}(Y, t_l; Y ,t_r)=\frac{N_{SL}}{2}
\frac{1}{(1+|\varepsilon|^2)^2}\;\frac{1}{1+|r_{00}|^2}\;\frac{1}{1+|r_{+-}|^2}}\nonumber\\
& &\biggl\lbrace|1+\varepsilon\;\varepsilon_{+-}^*|^2|\varepsilon_{00}^*+\varepsilon|^2
e^{-\Gamma_S t_l-\Gamma_L
t_r}+|\varepsilon_{+-}^*+\varepsilon|^2|1+\varepsilon\;\varepsilon_{00}^*|^2
e^{-\Gamma_L t_l-\Gamma_S t_r}\nonumber\\
& &-2 Re\{(1+\varepsilon^*\;\varepsilon_{+-})(\varepsilon_{00}+\varepsilon^*)
(\varepsilon_{+-}^*+\varepsilon)(1+\varepsilon\;\varepsilon_{00}^*)\;e^{-i\Delta m
\Delta t}\}\cdot e^{-\gamma (t_l+t_r)}\biggr\rbrace\\
\lefteqn{P_{K_{+-},K_{+-}}(Y, t_l; Y, t_r)=\frac{N_{SL}}{2}
\frac{1}{(1+|\varepsilon|^2)^2}\;\frac{1}{(1+|r_{+-}|^2)^2}\;
|1+\varepsilon\;\varepsilon_{+-}^*|^2\;|\varepsilon_{+-}^*+\varepsilon|^2}\nonumber\\
& &\biggl\lbrace
e^{-\Gamma_S t_l-\Gamma_L t_r}+e^{-\Gamma_L t_l-\Gamma_S t_r}-2 \cos(\Delta m \Delta
t)\cdot e^{-\Gamma (t_l+ t_r)}\biggr\rbrace
\end{eqnarray}

\newpage
The correction function:

\begin{eqnarray}
\lefteqn{h(K^0, t_a; K_{00}, t_b; K_{+-}, t_c; K_{+-}, t_d)=}\nonumber\\
& &-P_{K^0,K_{00}} (Y, t_a; Y, t_b)+P_{K^0, K_{+-}}(Y, t_a; Y
,t_c)+P_{K_{+-},K_{+-}}(Y,
t_d; Y, t_b)\nonumber\\
& &+P_{K_{+-},K_{00}}(Y, t_d; Y, t_c)\nonumber\\
& &+3-N_{SL}^2\biggl\lbrace
\frac{1+\delta}{2}\bigl(e^{-\Gamma_S t_a}+e^{-\Gamma_L t_a}-2 \delta \cos(\Delta m
t_a)\cdot e^{-\Gamma t_a}\bigr)\nonumber\\
& &\hphantom{+3-N_{SL}^2}+
\frac{1}{1+|\varepsilon|^2}\;\frac{1}{1+|\varepsilon_{+-}|^2}\bigl(|1+\varepsilon_{+-}^*\varepsilon|^2
e^{-\Gamma_S t_d}+|\varepsilon+\varepsilon_{+-}^*|^2 e^{-\Gamma_L t_d}\nonumber\\
& &\hphantom{+3-N_{SL}^2}
-2\delta Re\{(1+\varepsilon_{+-}\varepsilon^*)(\varepsilon
+\varepsilon_{+-}^*)e^{-i \Delta m t_d}\}\cdot e^{-\Gamma t_d}\bigr)\nonumber\\
& &\hphantom{+3-N_{SL}^2}+
\frac{1}{1+|\varepsilon|^2}\;\frac{1}{1+|\varepsilon_{00}|^2}\bigl(|1+\varepsilon_{00}^*\varepsilon|^2
e^{-\Gamma_S t_b}+|\varepsilon+\varepsilon_{00}^*|^2 e^{-\Gamma_L t_b}\nonumber\\
& &\hphantom{+3-N_{SL}^2}
-2\delta Re\{(1+\varepsilon_{00}\varepsilon^*)
(\varepsilon+\varepsilon_{00}^*)e^{-i \Delta m t_b}\}\cdot e^{-\Gamma t_b}\bigr)\nonumber\\
& &\hphantom{+3-N_{SL}^2}+
\frac{1}{1+|\varepsilon|^2}\;\frac{1}{1+|\varepsilon_{+-}|^2}\bigl(|1+\varepsilon_{+-}^*\varepsilon|^2
e^{-\Gamma_S t_c}+|\varepsilon+\varepsilon_{+-}^*|^2 e^{-\Gamma_L t_c}\nonumber\\
& &\hphantom{+3-N_{SL}^2}
-2\delta Re\{(1+\varepsilon_{+-}\varepsilon^*)(\varepsilon
+\varepsilon_{+-}^*)e^{-i \Delta m t_c}\}\cdot e^{-\Gamma t_c}\bigr)\biggr\rbrace
\end{eqnarray}

\end{document}